\documentclass[prl,twocolumn,showpacs,superscriptaddress,preprintnumbers,nofootinbib, amsmath,amssymb]{revtex4-1}
\usepackage{amsmath}
\usepackage{mathrsfs}
\usepackage{graphicx}
\usepackage{xcolor}
\usepackage{dcolumn}
\usepackage{bm}
\usepackage{amssymb}
\usepackage{amsmath}
\usepackage{verbatim}
\usepackage[utf8]{inputenc}
\usepackage{tikz,hyperref}
\usepackage{dcolumn}

\definecolor{darkyellow}{rgb}{0.545098,0,0}

\newcommand{\orcid}[1]{\href{https://orcid.org/#1}{\includegraphics[width=8pt]{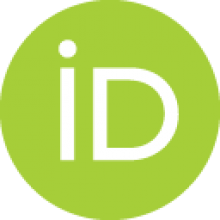}}}

\begin{document}

\title{Three Birds with One Stone: Core-Collapsed SIDM Halos as the Common Origin of Dense Perturbers in Lenses, Streams, and Satellites}

\author{Hai-Bo Yu\orcid{0000-0002-8421-8597}}
\email{haiboyu@ucr.edu}
\affiliation{Center for Experimental Cosmology \& Instrumentation, Department of Physics and Astronomy, University of California, Riverside, California 92521, USA}

\begin{abstract}

We show that core-collapsed self-interacting dark matter halos of mass $\sim 10^6\,{\rm M_\odot}$, originally simulated to explain the dense perturber of the GD-1 stellar stream, also reproduce the structural properties inferred for the dense perturber detected in the strong lensing system JVAS~B1938+666 from radio observations. Furthermore, these halos are sufficiently compact and dense to gravitationally capture field stars in satellite galaxies of the Milky Way, providing a natural explanation for the origin of Fornax~6, a stellar cluster in the Fornax dwarf spheroidal galaxy. Our results demonstrate that observations of halos with similar masses but residing in different cosmic environments offer a powerful and complementary probe of self-interacting dark matter.

\end{abstract}

\maketitle

\section{Introduction}

Recently, Ref.\,\cite{Powell:2025rmj} reported the discovery of an object that perturbs the lens JVAS~B1938+666 through gravitational imaging of radio observations. The detection has a high statistical significance of $26\sigma$ and a precisely measured mass of $(1.13\pm0.04)\times10^6\,{\rm M_\odot}$ within a projected radius of $80\,{\rm pc}$ at redshift $z=0.881$. The object can be well described by a truncated isothermal profile, $\rho(r)\propto r^{-2}$ toward the center, with a total mass of $\approx3\times10^6\,{\rm M_\odot}$, indicating that it is both dense and compact. This represents the lowest-mass object detected at a cosmological distance through its gravitational effect, and its density is exceptionally high.

Ref.\,\cite{Penarrubia:2024vms} showed that a dense object may exist within the Fornax satellite galaxy of the Milky Way. In addition to the five known globular clusters, Fornax has been confirmed to host a sixth stellar cluster, Fornax~6~\citep{DES:2019qxg}. The metallicity and age of the stars in Fornax~6 are similar to those of the metal-rich field stars in Fornax but distinct from those in the other five clusters. The inferred mass-to-light ratio is anomalously high $M/L\sim15\textup{--}258\,{\rm M_\odot/L_\odot}$~\citep{Pace:2021}, and no tidal tails are observed. Ref.\,\cite{Penarrubia:2024vms} proposed that Fornax~6 may have formed through the temporary capture of field stars by a dense substructure of mass $\sim10^6\,{\rm M_\odot}$ orbiting within the Fornax potential.

Furthermore, Ref.\,\cite{Bonaca:2018fek} reported the detection of a low-mass, dense perturber in the Milky Way through observations of the GD-1 stellar stream; see also~\cite{Nibauer:2025ezn}. The stream exhibits spur and gap features, indicating gravitational perturbation by an unseen object. To reproduce these features, particularly the spur-like distribution of stars displaced from the main stream, the perturber must be highly concentrated, with a mass in the range $10^{5.5}\textup{--}10^8\,{\rm M_\odot}$. {It is remarkable that these three objects, similar in mass, while residing in different environments, are all compact and dense. {If interpreted as dark matter halos, their inferred concentrations are systematically higher than the predictions of the cold dark matter (CDM) framework.}

In this work, we show that the structural properties of the B1938+666 strong-lensing perturber (radio), the Fornax substructure, and the GD-1 stream perturber are remarkably similar, suggesting a common physical origin despite their distinct astrophysical contexts. We further demonstrate that these properties can be naturally explained within the self-interacting dark matter (SIDM) framework, wherein dark matter halos undergo gravothermal collapse and develop dense, compact cores in their central regions (see~\cite{Tulin:2017ara-core,Adhikari:2022sbh} for reviews). In particular, we find that the core-collapsed SIDM halos simulated in~\cite{Zhang:2024fib}, originally proposed to account for the high density of the GD-1 perturber, also reproduce the structural properties inferred for both the B1938+666 perturber (radio) and the Fornax substructure.

\section{Density profiles}
\label{sec:cdm}

\begin{figure}[t]
	\centering
	\includegraphics[scale=0.45]{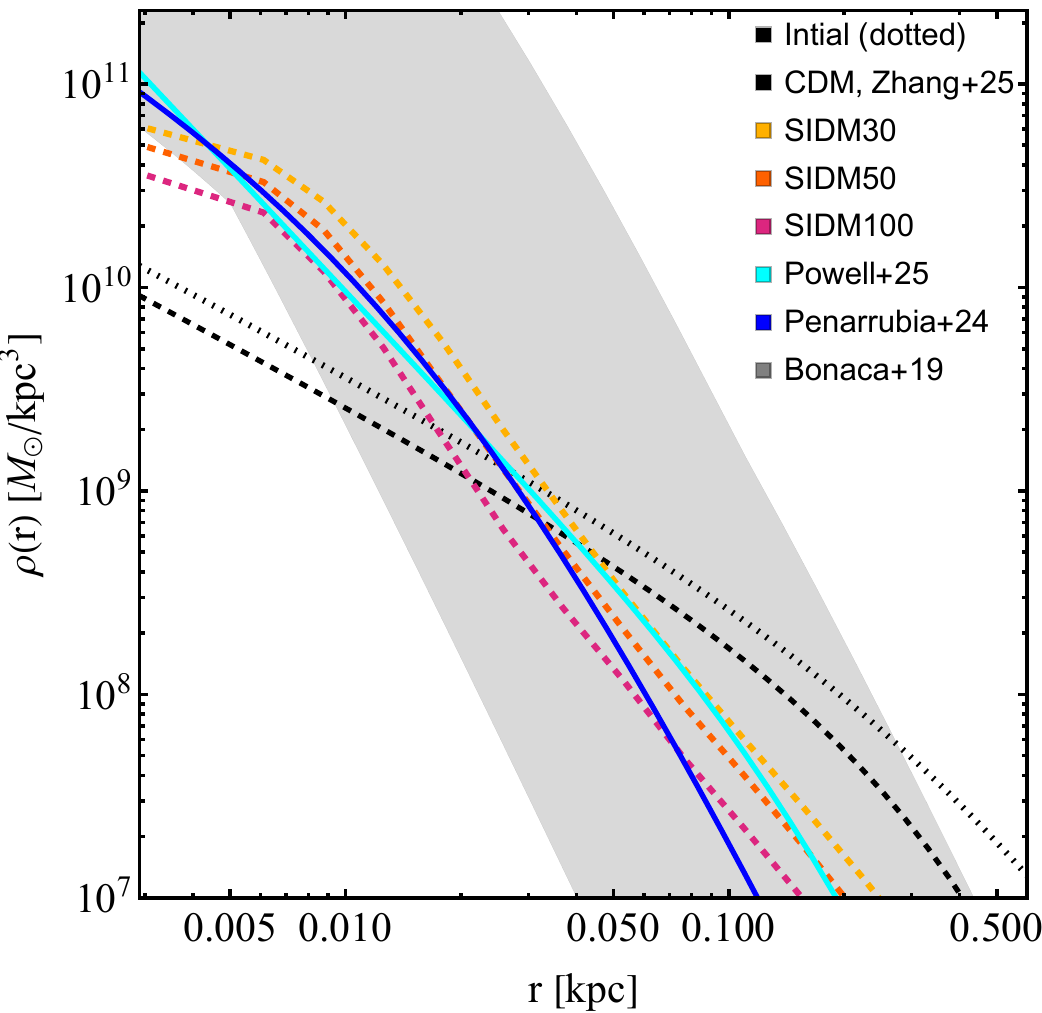}~
	\caption{Density profiles of the JVAS B1938+666 strong-lensing perturber from radio observations~\citep{Powell:2025rmj} (solid cyan), the Fornax substructure~\citep{Penarrubia:2024vms} (solid blue), and the GD-1 stream perturber~\citep{Bonaca:2018fek} (shaded gray band). For comparison, we show simulated halo density profiles for $\sigma/m=0\,{\rm cm^2/g}$ (CDM, dashed black), $30\,{\rm cm^2/g}$ (dashed amber), $50\,{\rm cm^2/g}$ (dashed orange), and $100\,{\rm cm^2/g}$ (dashed pink), as well as the initial condition before tidal evolution (dotted black), from~\cite{Zhang:2024fib}. {The legend indicates the color scheme of the curves and the band.}}
	\label{fig:combo}
\end{figure}

In this section, we compare density profiles of the three objects with characteristic masses of $\sim10^6\,{\rm M_\odot}$. For the B1938+666 perturber, we adopt the Pseudo-Jaffe profile used in~\cite{Powell:2025rmj}, given by
\begin{equation}
\label{eq:pj}
\rho(r) = \frac{\rho_0 r_t^4}{r^2(r^2 + r_t^2)},
\end{equation}
where $\rho_0$ denotes the normalization and $r_t$ is the truncation radius. Using the best-fit parameters reported in~\cite{Powell:2025rmj}, with a total mass of $(2.82 \pm 0.26)\times10^6\,{\rm M_\odot}$ and $r_t = (149 \pm 18)\,{\rm pc}$, we obtain $\rho_0 \approx 4.3\times10^7\,{\rm M_\odot/kpc^3}$ from Equation~\ref{eq:pj}. The corresponding density profile of the B1938+666 perturber is shown in Figure~\ref{fig:combo} (solid cyan).

For the Fornax substructure, we adopt a Hernquist profile~\citep{Hernquist:1990be}, following~\cite{Penarrubia:2024vms},
\begin{equation}
\label{eq:hq}
\rho(r) = \frac{M}{2\pi}\frac{a}{r}\frac{1}{(r + a)^3},
\end{equation}
where $M$ is the total mass and $a$ is the scale radius. Ref.\,\cite{Penarrubia:2024vms} proposed that the stellar cluster Fornax~6 may comprise field stars captured by a dense dark matter halo orbiting within the Fornax potential. {Fornax~6 has a stellar mass of $\approx7.2\times10^3\,{\rm M_\odot}$ and a half-light radius of $11\,{\rm pc}$~\citep{DES:2019qxg}. Its mean metallicity ${\rm [Fe/H]}\approx-0.71$ matches that of Fornax field stars but is much higher than in the other five clusters ($-2.5<{\rm [Fe/H]}<-1.4$)~\citep{Pace:2021}. It also shows a large velocity dispersion of $\approx5.6\,{\rm km/s}$, implying a mass-to-light ratio of $M/L\sim15\textup{--}258\,{\rm M_\odot/L_\odot}$~\citep{Pace:2021}. In the capture scenario, a substructure orbiting Fornax binds field stars. Reproducing Fornax~6 requires capturing $\sim10^4$ stars, achieving $M/L\sim100\,{\rm M_\odot/L_\odot}$, and producing a bound stellar density above the field background. Ref.\,\cite{Penarrubia:2024vms} showed this is feasible for a substructure with mass $\gtrsim10^6\,{\rm M_\odot}$ and scale radius $\lesssim20\,{\rm pc}$. For illustration, we adopt $M=10^6\,{\rm M_\odot}$ and $a=20\,{\rm pc}$; the profile is shown in Figure~\ref{fig:combo} (solid blue)}.

For the perturber of the GD-1 stellar stream, Ref.\,\cite{Bonaca:2018fek} modeled it using a Hernquist profile as given in Equation~\ref{eq:hq}, and inferred a parameter region corresponding to a mass of $10^{5.5}\textup{--}10^8\,{\rm M_\odot}$ and a scale length of $a\lesssim10\textup{--}30\,{\rm pc}$. Ref.\,\cite{Zhang:2024fib} converted this region into the corresponding range of density profiles for the perturber, which we show in Figure~\ref{fig:combo} (shaded gray). The density profile of the Fornax substructure (solid blue) serves as an excellent representative example for the GD-1 perturber; therefore, we do not present a separate specific example for the latter.

We compare the density profiles of the three objects with those of the simulated halos presented in~\cite{Zhang:2024fib}, which were designed to investigate whether the high density of the GD-1 perturber can be explained by SIDM core collapse. The simulations were performed for a Milky Way–like system with a static host potential that includes both dark matter and baryonic components. The initial condition of the halo corresponds to a progenitor immediately prior to infall, extracted from a cosmological zoom-in CDM simulation of a Milky Way analog~\citep{Yang:2022mxl-tid-cosmo-CDM}; see Figure~\ref{fig:combo} (dotted black). {The initial halo follows a Navarro-Frenk-White (NFW) profile~\citep{Navarro:1996gj} with the scale density and radius $\rho_s\approx7.5\times10^7\,{\rm M_\odot/kpc^3}$ and $r_s\approx0.50\,{\rm kpc}$, respectively. Its concentration is $28$, which is $1.5\sigma$ higher than the cosmological median ($z=0$)~\citep{Dutton:2014xda}. The simulation particle mass is $32.5\,{\rm M_\odot}$ with a total number of $10^7$. The softening length is $\epsilon=2\,{\rm pc}$, resulting in a spline length of $\ell=2.8\epsilon\approx6\,{\rm pc}$.}       

Ref.\,\cite{Zhang:2024fib} evolved the halo in the Galactic tidal field assuming SIDM cross sections per unit mass of $\sigma/m=0\,{\rm cm^2/g}$ (CDM), $30\,{\rm cm^2/g}$, $50\,{\rm cm^2/g}$, and $100\,{\rm cm^2/g}$. The corresponding density profiles are shown in Figure~\ref{fig:combo} as dashed black, amber, orange, and pink curves, respectively,  {from the snapshots at $t=10\,{\rm Gyr}$}. The initial halo mass is $\approx3.3\times10^{8}\,{\rm M_\odot}$ ({$V_{\rm max}\approx15\,{\rm km/s}$}), and the final bound mass ranges from $4\times10^6\,{\rm M_\odot}$ to $10^7\,{\rm M_\odot}$ in the SIDM runs, with larger cross sections leading to smaller final masses due to the kick-out effect~\citep{Kong:2025kkt}. {The simulation follows the full gravothermal evolution of the halo. It first enters a core-expansion phase, where the inner density decreases drastically, and then transitions to the core-collapse phase at $t\sim2\,{\rm Gyr}$, after which the central density increases. The central density evolution largely saturates after $t\sim4\textup{--}6\,{\rm Gyr}$ due to limited resolution in the deep core phase, which prevents resolving the innermost collapse; the profiles shown here are therefore likely conservative; see~\cite{Zhang:2024fib}.}

From Figure~\ref{fig:combo}, we find that the inferred density profiles of the B1938+666 perturber, the Fornax substructure, and the GD-1 perturber are remarkably similar. All three are systematically denser and more compact in their inner regions than expected in the CDM framework, but they align closely with the profiles of core-collapsed SIDM halos. Despite residing in distinct environments, SIDM provides a unified explanation for their high densities, with a cross section of $\sigma/m = 30\textup{--}100\,{\rm cm^2/g}$. {We note that the SIDM simulations are not intended to reproduce exactly the density profiles of the observed dense objects, particularly in the outer regions, nor their total masses. As discussed above, current observational constraints on the GD-1 perturber and the Fornax substructure remain broad, as illustrated by the shaded band in Figure~\ref{fig:combo} (GD-1); see also~\cite{Bonaca:2018fek,Penarrubia:2024vms}. The simulated SIDM profiles lie well within these favored regions. The B1938+666 perturber is more tightly constrained, with uncertainties of $\sim10\%$ for a given density profile~\citep{Powell:2025rmj}. A dedicated strong-lensing analysis based directly on SIDM-motivated profiles would provide a more decisive test of the core-collapse interpretation.}

{The initial mass of the simulated SIDM halos ($3.3\times10^8\,{\rm M_\odot}$) is close to the galaxy formation threshold~\citep{Manwadkar2022,Ahvazi2024,Nadler:2025mnz,Benitez-Llambay:2020zbo}. The progenitors of the dense objects could have hosted very faint galaxies. However, any surviving stellar component at present is expected to be small or negligible due to severe tidal stripping. More broadly, the amount of stellar remnant depends on the initial stellar mass and distribution, orbital history, and the nature of dark matter. Compared to CDM, SIDM halos experience enhanced tidal mass loss from intermediate to outer regions (e.g.,~\cite{Kong:2025kkt}), leading to greater stellar stripping. Among the three objects considered, the B1938+666 perturber resides in a more massive host halo (a few times $10^{13}\,{\rm M_\odot}$), implying a potentially more massive progenitor. It would be interesting to perform simulations including live stellar particles to further investigate the contribution of stellar mass in this system.}

\section{Massive CDM halos on the tidal track}

    \begin{figure}[h]
    	\centering
		\includegraphics[scale=0.5]{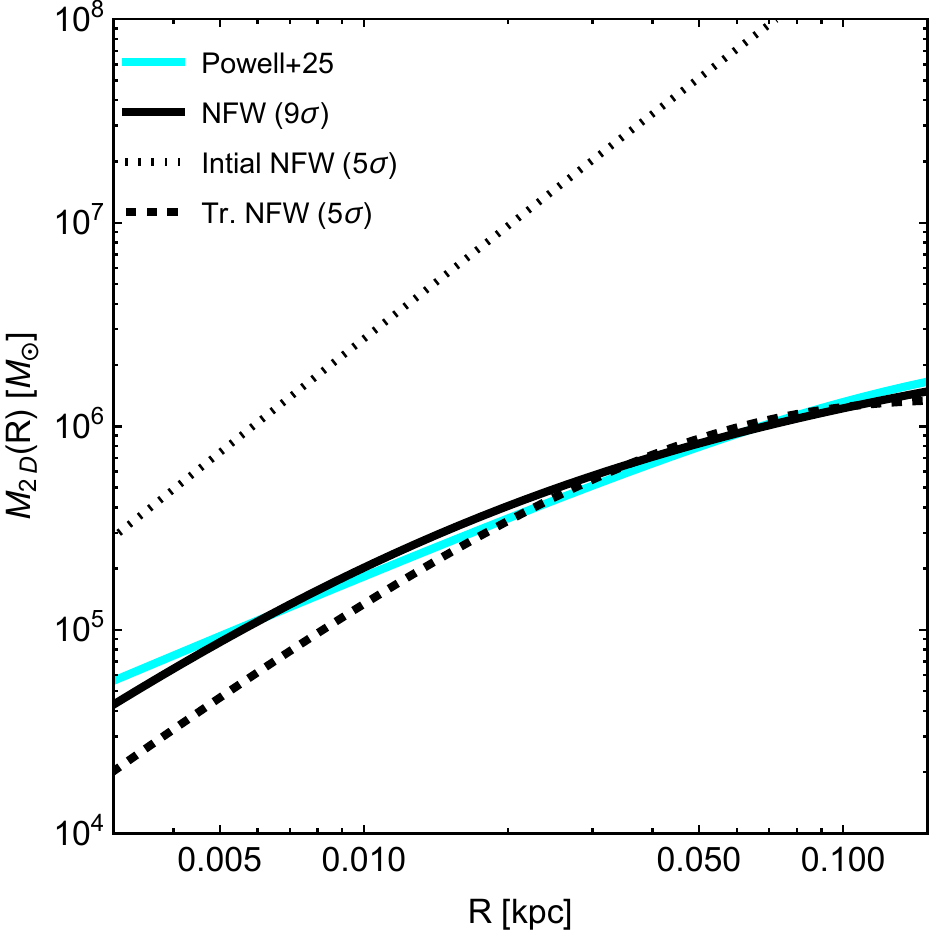}
		\caption{{The cylindrical mass profiles for the B1938+666 strong-lensing perturber~\citep{Powell:2025rmj} (solid cyan), an NFW halo (solid black) with mass $M_{200}=3.5\times10^6\,{\rm M_\odot}$ and concentration $c_{200}\approx291$, which is $9\sigma$ above the cosmological median for isolated halos, and a truncated NFW halo (dashed black), together with its initial halo (dotted black) with $M_{200}=10^{12}\,{\rm M_\odot}$ and concentration $c_{200}\approx30$, corresponding to $5\sigma$ above the median.}}
		\label{fig:m2d}
    \end{figure}

{Although Figure~\ref{fig:combo} shows only one CDM halo, Ref.\,\cite{Zhang:2024fib} analyzed $125$ progenitor halos with masses $10^8\textup{--}10^{10}\,{\rm M_\odot}$ in a Milky Way–like system from cosmological zoom-in simulations~\citep{Yang:2022mxl-tid-cosmo-CDM} and found none dense enough to match the GD-1 perturber. In addition, using the subhalo concentration–mass relation derived in~\cite{Moline:2016pbm}, Ref.\,\cite{Bonaca:2018fek} showed that a GD-1 perturber model satisfying the minimum density requirement can be consistent with CDM at the $3\sigma$ level. The Fornax substructure is expected to follow a similar trend~\citep{Penarrubia:2024vms}. A more thorough assessment of CDM predictions requires a large halo sample in their respective environments, which we leave for future work. }

{We next test whether the high density of the B1938+666 perturber can be accommodated in CDM for a broader progenitor mass range using the tidal-track model~\citep{Errani:2020wgn}, where the density profile follows}
{
\begin{equation}
\rho(r)=\frac{\rho_s}{(r/r_s)(1+r/r_s)^2}\times\frac{\exp(-r/r_{\rm cut})}{(1+r_s/r_{\rm cut})^{0.3}},
\label{eq:tidal}
\end{equation}
with scale parameters $r_s$, $\rho_s$, and cutoff radius $r_{\rm cut}$. We also calculate the cylindrical mass profile with the projected radius $R$ as} 

{
\begin{equation}
M_{\rm 2D}(R)=2\pi\int^R_0 R'dR'\int dz \rho(\sqrt{R'^2+z^2})
\end{equation}}

{We consider two extremes. First, taking $r_s/r_{\rm cut}\rightarrow0$, a low-mass NFW halo matching the perturber density requires $M_{200}=3.5\times10^6\,{\rm M_\odot}$ and $c_{200}\approx291$, $9\sigma$ above the cosmological median for isolated halos ($z=0$)~\citep{Dutton:2014xda}, corresponding to $\rho_s\approx4.4\times10^{10}\,{\rm M_\odot/kpc^3}$ and $r_s\approx0.011\,{\rm kpc}$. Figure~\ref{fig:m2d} shows that this halo (solid black) reproduces the cylindrical mass profile of the perturber model (solid cyan). For the B1938+666 perturber, the characteristic projected radius is $R=80\,{\rm pc}$ and the mass is $M_{\rm 2D}(80\,{\rm pc})=(1.13\pm0.04)\times10^6\,{\rm M_\odot}$~\citep{Powell:2025rmj}.}

{Second, we consider an initial massive halo with $M_{200}=10^{12}\,{\rm M_\odot}$ and $c_{200}\approx30$, $5\sigma$ above the median for isolated halos, corresponding to $\rho_s\approx8.9\times10^7\,{\rm M_\odot/kpc^3}$ and $r_s\approx7.1\,{\rm kpc}$. Assuming the halo follows the tidal track, we model its subsequent density profile using Equation~\ref{eq:tidal} with $r_{\rm cut}=0.03\,{\rm kpc}$. Figure~\ref{fig:m2d} shows the truncated halo (dashed black) and its initial condition (dotted black). The truncated halo yields a cylindrical mass profile comparable to the B1938+666 perturber model for $R=10\textup{--}80\,{\rm pc}$. The adopted $r_{\rm cut}=0.03\,{\rm kpc}$ is consistent with the tidal-track value set by the ratio of initial to final masses (Equation 9 of~\cite{Errani:2020wgn}).}

{We see that the required concentration decreases as the progenitor mass increases. However, even for a $10^{12}\,{\rm M_\odot}$ progenitor, it would remain a $5\sigma$ outlier in the CDM concentration distribution. Moreover, this massive-progenitor scenario does not yield a sufficiently high central density for $R\lesssim10\,{\rm pc}$, although this region is not well constrained by current observations. We therefore conclude that interpreting the B1938+666 perturber as a CDM subhalo is challenging, unless a substantial stellar component contributes, as discussed in the previous section.}

Lastly, we cannot rule out the possibility that the B1938+666 and GD-1 perturbers are unidentified globular clusters. Indeed, core-collapsed SIDM halos and globular clusters can exhibit similar inner structures, both of which can be described by King profiles~\citep{Zhang:2024fib,Fischer:2025rky}. Future observations will be crucial to distinguish between these two scenarios. Furthermore, more precise measurements of the Fornax satellite and its stellar cluster Fornax~6 could provide a valuable test of the capture mechanism proposed in~\cite{Penarrubia:2024vms}.

\section{Discussions and conclusion}
\label{sec:diss}

{In this section, we motivate the cross section range adopted in this work and relate it to velocity-dependent SIDM models studied in recent cosmological zoom-in simulations. These models may offer a unified core-collapse interpretation of dense, compact (sub)structures in strong-lensing systems across a broad range of masses. We also outline directions for further improving this analysis. } 

{Ref.\,\cite{Zhang:2024fib} assumed constant self-interaction cross sections $\sigma/m=30\textup{--}100\,{\rm cm^2/g}$ for halo masses $10^6\textup{--}10^8\,{\rm M_\odot}$. For an isolated $10^8\,{\rm M_\odot}$ dwarf halo ($V_{\rm max}\sim10\,{\rm km/s}$) with a concentration $2\sigma$ above the cosmological median~\citep{Dutton:2014xda}, the core-collapse timescale is shorter than a Hubble time for $\sigma/m\gtrsim30\,{\rm cm^2/g}$ (Eq.~2 of~\cite{Zhang:2024fib}), setting the lower bound. Velocity-dependent SIDM models favor larger amplitudes up to $\sim100\,{\rm cm^2/g}$ to induce core collapse in more massive (sub)structures. This is realized in the SIDM Concerto suite~\citep{Nadler:2023nrd-tid,Nadler:2025jwh}, where the GroupSIDM-70 and GroupSIDM-147 models asymptote to $70\,{\rm cm^2/g}$ and $147\,{\rm cm^2/g}$ below $10^8\,{\rm M_\odot}$. \cite{Kong:2025sqx} showed that core-collapsed halos from these simulations can reproduce the high densities inferred for more massive strong-lensing objects, including the J0946+1006 perturber ($10^{10}\,{\rm M_\odot}$)~\citep{Vegetti:2009cz,Minor:2020hic,Enzi:2024ygw,Despali:2024ihn,Minor:2025}, and the B1938+666 perturber inferred from optical observations ($5\times10^8\,{\rm M_\odot}$)~\citep{Vegetti:2012mc,Sengul:2022edu,Despali:2024ihn,Tajalli:2025qjx,Lei:2025pky}.}

{Therefore, the mass scale of dense objects inferred from observations is not limited to $10^6\,{\rm M_\odot}$, the focus of this work. Their population may follow a broader distribution, though the current sample is too small to confirm this. In velocity-dependent SIDM models, where the cross section increases as halo mass decreases, there exists an optimal mass scale at which the core-collapse fraction peaks~\citep{Ando:2024kpk,Nadler:2025jwh,Jiang:2025jtr}. For example, in the GroupSIDM-147 model, the peak velocities of the optimal mass scales are $30\,{\rm km/s}$ and $60\,{\rm km/s}$ for isolated halos and subhalos, respectively (e.g, Figure 5 of~\cite{Nadler:2025jwh}). Future work with larger samples of dense objects, combined with forward modeling and careful treatment of observational biases, will allow more robust constraints. Interestingly, similar velocity-dependent SIDM models have been proposed to explain the formation of supermassive black holes in the early Universe via core collapse (e.g.,~\cite{Jiang:2025jtr,Feng:2025rzf,Feng:2020kxv}). }

While the inferred concentration of the higher-mass J0946+1006 perturber remains uncertain due to possible contamination from luminous matter~\citep{Li:2025kpb,He:2025wco}, such contamination effects are expected to be negligible for both B1938+666 perturbers because of their low masses. These systems therefore offer particularly clean and critical tests of the SIDM interpretation. We expect that the Concerto suite also contains core-collapsed halos with masses of order $10^6\,{\rm M_\odot}$, but such systems are not well resolved. The particle mass in the highest-resolution simulation, corresponding to an LMC-like system, is $6.3\times10^{3}\,{\rm M_\odot}$, while it increases to $4.0\times10^{5}\,{\rm M_\odot}$ for a group-scale system~\citep{Nadler:2025jwh}. Thus, constructing reliable inner density profiles for $\sim10^6\,{\rm M_\odot}$ halos remains challenging, even in the LMC analog runs.

On the other hand, the simulations shown in Figure~\ref{fig:combo} use an idealized setup with a particle mass of $32.5\,{\rm M_\odot}$~\citep{Zhang:2024fib}. These simulations do not account for the realistic tidal environments and assembly histories of the B1938+666 perturber and the Fornax substructure. Nevertheless, our main conclusion that core-collapsed halos can reproduce their high densities remains robust because of the self-similar nature of SIDM halo evolution~\citep{Outmezguine:2022bhq,Zhong:2023yzk,Fischer:2025rky}. Moreover, Ref.\,\cite{Kong:2025sqx} found that effective concentrations of SIDM halos in the Concerto suite do not exhibit biases across different host mass scales, further supporting our conclusion.

To further break degeneracies among model parameters, such as halo concentration, tidal orbit, and scattering cross section, system-specific simulations will be required. A promising direction is to use a hybrid simulation approach as in~\cite{Zhang:2024fib,Zhang:2024ggu}, where the initial halo conditions are drawn from cosmological simulations (e.g., the Concerto suite), while the host is modeled using a semi-analytic potential. This approach would enable ultra-high-resolution studies at modest computational cost. We leave such investigations for future work.

In summary, we have shown that the SIDM core-collapse scenario can naturally account for the high densities of three $\sim10^6\,{\rm M_\odot}$ objects inferred from observations of the strong lens B1938+666, the Fornax satellite galaxy, and the GD-1 stellar stream. This unified explanation connects three distinct observational systems within a single physical framework. The preferred range of SIDM cross sections is consistent with velocity-dependent models previously proposed to explain the high densities of more massive substructures. In the future, system-specific simulations and modeling will be essential to further break parameter degeneracies, and improved observational constraints will be critical for reducing astrophysical uncertainties. Together, these efforts will provide decisive tests of the new-physics interpretation of these intriguing ``outliers.''

\section*{acknowledgments}
We thank Demao Kong for helpful discussions and Xingyu Zhang for providing the simulation data. This research was supported by the John Templeton Foundation under Grant ID\#61884 and the U.S. Department of Energy under Grant No.~DE-SC0008541. The opinions expressed in this publication are those of the authors and do not necessarily reflect the views of the funding agencies.

\section*{DATA AVAILABILITY}

The Mathematica notebook used to produce the plots in Figures~\ref{fig:combo} and~\ref{fig:m2d} is available at \cite{data:2026}.

\bibliography{refbib-perturber}

\end{document}